\documentclass[aps,pra,showpacs,twocolumn]{revtex4}

\usepackage{dcolumn}
\usepackage{bm}
\usepackage{amsmath}
\usepackage{graphicx}
\usepackage{amsmath,amssymb}

\newcommand{\mean}[1]{\langle{#1}\rangle}

\begin{document}

\preprint{APS/123-QED}

\title{
Quantum Feedback Amplification
}

\author{Naoki Yamamoto}
\affiliation{%
Department of Applied Physics and Physico-Informatics,
Keio University, 
Hiyoshi 3-14-1, Kohoku, Yokohama 223-8522, Japan
}

\date{\today}

\begin{abstract}

Quantum amplification is essential for various quantum technologies such as 
communication and weak-signal detection. 
However, its practical use is still limited due to inevitable device fragility that 
brings about distortion in the output signal or state. 
This paper presents a general theory that solves this critical issue. 
The key idea is simple and easy to implement: 
just a passive feedback of the amplifier's auxiliary mode, which is usually 
thrown away. 
In fact, this scheme makes the controlled amplifier significantly robust, and 
furthermore it realizes the minimum-noise amplification even under realistic 
imperfections. 
Hence, the presented theory enables the quantum amplification to be 
implemented at a practical level. 
Also, a nondegenerate parametric amplifier subjected to a special detuning 
is proposed to show that, additionally, it has a broadband nature. 

\end{abstract}

\pacs{42.65.Yj, 02.30.Yy, 03.65.Yz, 42.50.Lc}
\maketitle


\section{Introduction}

The amplifier is clearly one of the most important components incorporated in 
almost all current technological devices. 
The basic function of an autonomous amplifier is simply to transform an input 
signal $u$ to $y=Gu$ with gain $G>1$. 
However, such an amplifier is fragile in the sense that the device parameters 
change easily, and eventually distortion occurs in the output $y$. 
This was indeed a most serious issue which had prevented any practical use 
of amplifiers in, e.g., telecommunication. 
Fortunately, this issue was finally resolved back in 1927 by Black 
\cite{Black 77,Black 84}; 
there are a huge number of textbooks and articles reviewing this revolutionary 
work, and here we refer to Refs. \cite{Mancini,Bechhoefer}. 
The key idea is the use of feedback shown in Fig.~1; 
that is, an autonomous amplifier called the ``plant" is combined with a ``controller" 
in such a way that a portion of the plant's output is fed back to the plant through 
the controller. 
Then the output of the whole controlled system is given by 
\begin{equation}
\label{classical FB}
     y=G^{(\rm fb)}u,~~~G^{({\rm fb})}=\frac{G}{1+GK}, 
\end{equation}
where $K$ is the gain of the controller. 
Now, if the plant has a large gain $G\gg 1$, it immediately follows that 
$G^{({\rm fb})}\approx 1/K$. 
Hence, the whole system works as an amplifier, simply provided that the controller 
is a passive device (i.e., an attenuator) with $K < 1$. 
Importantly, a passive device such as a resistor is very robust, and its parameters 
contained in $K$ almost do not change. 
This is the mechanism of robust amplification realized by feedback control. 
Note, of course, that this feedback architecture is the core of an operational 
amplifier (op-amp).

\begin{figure}[t]
\centering 
\includegraphics[width=8.5cm]{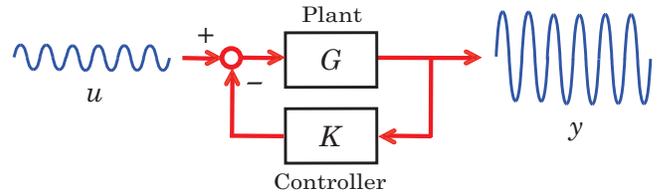}
\caption{
Classical feedback-amplification scheme: $G$ is the gain of an autonomous 
amplifier, and $K$ is the gain of a passive controller. 
}
\label{examples}
\end{figure}

Surely there is no doubt about the importance of quantum amplifiers. 
A pertinent quantum counterpart to the classical amplifier is the 
{\it phase-preserving linear amplifier} \cite{Haus 62,Caves 82} (in what 
follows, we simply call it the ``amplifier"). 
In fact, this system has a crucial role in diverse quantum technologies such as 
communication, weak-signal detection, and state processing 
\cite{Braunstein 01,Loudon 03,Leuchs 2008,Clerk 10,Hudelist 14,Ping Koy 14,
Lehnert 14,Girvin 15}. 
In particular, recent substantial progress in both theory and experiments 
\cite{Josse 06, Devoret 10,Devoret 10b,Nha 10,Furusawa 11,Caves 12,
Clerk 14,Abdo 14,Clerk 15,Mabuchi 15} has further advanced this field. 
An important fact is that, however, an amplifier must be an active system 
powered by external energy sources, implying that its parameters are 
fragile and can change easily. 
Because of this parameter fluctuation, the amplified output signal or state 
suffers from distortion \cite{Slavik 2010,Caves 13,Dastjerdi 13}. 
As a consequence, the practical applicability of the quantum amplification is 
still severely limited. 
That is, we are now facing the same problem we had 90 years ago.

To make the discussion clear, let us here describe the general 
quantum-amplification process. 
Ideally the amplifier transforms a bosonic input mode $b_1$ to 
$\tilde{b}_1 = g_1 b_1 + g_2 b_2^\dagger$, where $b_2$ is an auxiliary 
mode, and the coefficients satisfy $|g_1|^2 - |g_2|^2 = 1$ from 
$[\tilde{b}_1, \tilde{b}_1^\dagger]=1$. 
Hence, the output $\tilde{b}_1$ is an amplified mode of $b_1$ with gain 
$|g_1|>1$. 
A typical example of an amplifier is the optical nondegenerate parametric 
amplifier (NDPA), in which case $g_1$ and $g_2$ are frequency dependent 
as shown later. 
However, note again that the system parameters, especially the coupling 
strength of the pumped crystal, cannot be kept exactly constant, and 
eventually the amplified output mode $\tilde{b}_1$ has to be distorted.

Now the motivation is clear; we need a quantum version of the 
feedback-amplification method described in the first paragraph. 
The contribution of this paper is, in fact, to develop a general theory for 
quantum feedback amplification that resolves the fragility issue of quantum 
amplifiers. 
The key idea is simple and easy to implement, i.e., {\it feedback of the auxiliary 
output mode $\tilde{b}_2$ through a passive controller to the auxiliary 
input mode $b_2$}. 
Indeed, it is proven that the whole controlled system possesses a strong 
robustness property against parameter fluctuations, which thus enables 
quantum amplifiers to be implemented at a practical level. 
This type of control scheme is, in general, called the {\it coherent feedback} 
\cite{Wiseman 94,James 08,Gough 09,Mabuchi 13,Kerckhoff 13,Yamamoto 14}, 
meaning that an output field is fed back to an input field through another 
quantum system without involving any measurement process; 
hence, an excess classical noise is not introduced in the feedback loop. 
Now note that the auxiliary output $\tilde{b}_2$ has some information about 
$\tilde{b}_1$ due to their entanglement, though $\tilde{b}_2$ is usually 
thrown away in the scenario of quantum amplification. 
Thus, we have an interpretation that the presented scheme utilizes the 
{\it signal-recycling} technique \cite{Yanbei 02} for reducing the sensitivity, 
unlike the conventional use of it for enhancing the sensitivity of the 
gravitational-wave detector.

In addition to the above-described main contribution, some important results 
are obtained. 
First, we see that the controlled system reaches the fundamental 
quantum noise limit \cite{Caves 82} even if some imperfections are present 
in the feedback loop. 
This means that precise fabrication of the feedback control is not necessary, 
which thus again emphasizes the feasibility of the presented scheme. 
Next, this paper proposes a type of NDPA subjected to a special detuning 
that circumvents the usual gain-bandwidth trade-off in the amplification 
process. 
A drawback of this modified amplifier is that, as will be shown, it is very 
sensitive to the parameter fluctuation. 
The presented theory has a distinct advantage in such a situation; 
that is, this issue can now be resolved by constructing a feedback loop. 
Therefore, as a concrete application of the theory, this paper proposes a 
robust, near-minimum-noise, and broadband amplifier.

Finally, note that there are a variety of quantum amplifiers considered in the 
literature such as an optical back-action evasion amplifier \cite{Yurke}; 
however, the schematic presented in this paper is essentially different from 
all those modifications in the following sense. 
While those modified amplifiers have their own purposes for improving the 
performance or achieving the goal in some specific subjects 
(e.g. back-action evasion), the feedback scheme is a device-independent and 
purpose-independent fundamental architecture that must be incorporated in 
all amplifiers. 
In fact, in the classical regime, the ``operation" part of an op-amp has its 
own purpose (e.g., differentiation and integration), but any op-amp does not 
work without feedback.


\section{Model of phase-preserving linear quantum amplifier}

Let us begin with a specific model: the NDPA. 
This is an optical cavity system with two internal modes $a_1$ and $a_2$. 
They are orthogonally polarized and obey the following Hamiltonian: 
\[
      H= \omega_1 a_1^\dagger a_1 + \omega_2 a_2^\dagger a_2 
                + i\lambda (a_1^\dagger a_2^\dagger e^{-2i\omega_0 t} 
                                     - a_1 a_2 e^{2i\omega_0 t}),
\]
with $\lambda\in{\mathbb R}$ the coupling strength between the modes, 
$\omega_i$ the resonant frequencies of $a_i$, and $2\omega_0$ the pump 
frequency. 
Also, in the above expression the rotating-wave approximation is taken under 
the assumption $2\omega_0\approx \omega_1 + \omega_2$. 
The system couples with a signal input $b_1$ and an auxiliary (idler) input $b_2$ 
with strength $\kappa$. 
Then, in the rotating frame at frequency $\omega_0$, the dynamics of the NDPA 
is given by the following Langevin equations \cite{Clerk 10,Ou,Gardiner book}: 
\begin{eqnarray}
& & \hspace*{-1em}
\label{NDPA dynamics 1}
     \frac{da_1}{dt} 
              = \Big(-\frac{\kappa}{2}-i\Delta_1\Big)a_1 + \lambda a_2^\dagger 
                          -\sqrt{\kappa}b_1,
\\ & & \hspace*{-1em}
\label{NDPA dynamics 2}
     \frac{da_2^\dagger}{dt} 
              = \Big(-\frac{\kappa}{2}+i\Delta_2\Big)a_2^\dagger + \lambda a_1 
                          -\sqrt{\kappa}b_2^\dagger,
\end{eqnarray}
where $\Delta_1=\omega_1-\omega_0$ and $\Delta_2=\omega_2-\omega_0$ 
are detuning. 
Also, the output equations (boundary conditions) are given by 
\begin{equation}
\label{NDPA output}
     \tilde{b}_1=\sqrt{\kappa}a_1 + b_1,~~~
     \tilde{b}_2^\dagger=\sqrt{\kappa}a_2^\dagger + b_2^\dagger. 
\end{equation}
Now the Laplace transformation of an observable $x_t$ in the Heisenberg picture 
is defined by 
\[
    x(s) := \int_0^\infty e^{-st} x_tdt, 
\]
where ${\rm Re}(s)>0$. 
Then the Laplace transforms of $b_1$, etc., are connected by the following 
linear equations: 
\begin{eqnarray}
& & \hspace*{-1.2em}
     \tilde{b}_1(s) 
      = g_1(s) b_1(s) + g_2(s) b_2^\dagger(s), 
\nonumber \\ & & \hspace*{-1.2em}
      g_1(s)  = \frac{(s-\frac{\kappa}{2}+i\Delta_1)
                              (s+\frac{\kappa}{2}-i\Delta_2)-\lambda^2}{D(s)},~
      g_2(s)  =  \frac{-\kappa \lambda}{D(s)}, 
\nonumber \\ & & \hspace*{-1.2em}
      D(s) = \Big(s+\frac{\kappa}{2}+i\Delta_1\Big)
                 \Big(s+\frac{\kappa}{2}-i\Delta_2\Big) - \lambda^2.
\nonumber
\end{eqnarray}
The stability analysis can be conducted in the Laplace domain; 
that is, for the amplifier to be stable, all roots of the characteristic equations 
of the transfer functions (i.e., {\it poles}) must lie in the left-hand complex plane. 
In the above case, particularly when $\Delta_1=\Delta_2=0$, the characteristic 
equation is $D(s)=s^2+\kappa s +\kappa^2/4-\lambda^2=0$; 
hence, $\kappa^2/4-\lambda^2>0$ must be satisfied to guarantee the stability 
of the NDPA.

The quantum-amplification process is described in the Fourier domain 
$s=i\omega$ with $\omega$ the frequency; that is, we consider the linear 
transformation at the steady state, 
$\tilde{b}_1(i\omega) = g_1(i\omega) b_1(i\omega) 
+ g_2(i\omega) b_2^\dagger(i\omega)$. 
Note that $g_1$ and $g_2$ satisfy $|g_1(i\omega)|^2-|g_2(i\omega)|^2=1$ 
for all $\omega$. 
In particular, when $\Delta_1=\Delta_2=0$, the amplification gain at the 
resonant frequency $\omega=0$ is given by 
\[
        |g_1(0)|=\frac{\kappa^2 + 4\lambda^2}{|\kappa^2 - 4\lambda^2|},
\]
and it takes a large number nearly at the threshold 
$\lambda\approx \kappa/2-0$. 
Thus, $\tilde{b}_1$ is in fact an amplified mode of $b_1$ with gain $|g_1|$.

The above example can be generalized; 
any phase-preserving linear quantum amplifier is modeled as an open dynamical 
system with two inputs and two outputs. 
Let us represent the input-output relation in the Laplace domain as follows: 
\begin{eqnarray}
& & \hspace*{0em}
\label{system}
      \left[ \begin{array}{c}
                \tilde{b}_1(s) \\ 
                \tilde{b}_2^\dagger(s) \\
            \end{array} \right]
      = G(s) \left[ \begin{array}{c}
                     b_1(s) \\ 
                     b_2^\dagger(s) \\
                 \end{array} \right],
\nonumber \\ & & \hspace*{1.7em}
       G(s) 
         = \left[ \begin{array}{cc}
             G_{11}(s) & G_{12}(s) \\
             G_{21}(s) & G_{22}(s) \\
            \end{array} \right], 
\end{eqnarray}
where $b_1(s)$ is the Laplace transformation of $b_1$, etc. 
The transfer function matrix $G(s)$ at $s=i\omega$ (i.e. the scattering matrix) 
satisfies 
\begin{eqnarray}
& & \hspace*{-1em}
\label{G conditions}
    |G_{11}(i\omega)|^2 - |G_{12}(i\omega)|^2 =
    |G_{22}(i\omega)|^2 - |G_{21}(i\omega)|^2 = 1,
\nonumber \\ & & \hspace*{-1em}
    G_{21}(i\omega)G_{11}^*(i\omega) - G_{22}(i\omega)G_{12}^*(i\omega)=0~~~
    \forall \omega. 
\end{eqnarray}
Thus, $|G_{11}(i\omega)|$ represents the amplification gain.


\section{The quantum feedback amplification}

\subsection{Feedback configuration}

\begin{figure}[t]
\centering 
\includegraphics[width=7.2cm]{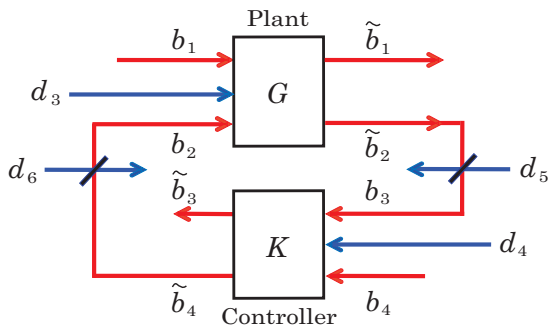}
\caption{
Coherent feedback configuration for the autonomous amplifier $G$ (plant). 
$d_3, d_4, d_5$, and $d_6$ are unwanted noisy input fields. 
}
\label{examples}
\end{figure}

Our control scheme is based on coherent feedback; 
that is, the controller is also given by a quantum system and is connected to 
the plant through the input and output fields. 
Note that, if a measurement process is involved in the feedback loop, it 
inevitably introduces additional noise. 
Now we take a passive system (e.g. a beam splitter and an optical cavity) 
as the controller, with two inputs $b_3, b_4$ and two outputs 
$\tilde{b}_3, \tilde{b}_4$; 
note that a single-input and single-output passive system has a gain equal to 1 
and thus does not work as an attenuator. 
We represent the input-output relation of this system in the Laplace domain 
as follows: 
\begin{eqnarray}
& & \hspace*{0em}
\label{controller}
      \left[ \begin{array}{c}
                \tilde{b}_3^\dagger(s) \\ 
                \tilde{b}_4^\dagger(s) \\
            \end{array} \right]
       = K(s)  \left[ \begin{array}{c}
                     b_3^\dagger(s) \\ 
                     b_4^\dagger(s) \\
                   \end{array} \right],
\nonumber \\ & & \hspace*{1.7em}
       K(s) = \left[ \begin{array}{cc}
                     K_{11}(s) & K_{12}(s) \\
                     K_{21}(s) & K_{22}(s) \\
                   \end{array} \right].
\end{eqnarray}
Here, the creation operator representation is taken to make the notation simple. 
Because of the passivity property, the transfer function matrix $K(s)$ is unitary 
in the Fourier domain; 
i.e., $K(i\omega)^\dagger K(i\omega)=I$ holds for all $\omega$.

We now consider connecting the controller to the plant. 
But unlike the classical case, where both the plant and the controller can be 
a single input-output system and arbitrary split or addition of signal is allowed, 
designing a feedback scheme in the quantum case is not trivial. 
For example, we could divide $\tilde{b}_1$ into two paths by a beam splitter 
and use one of them for feedback purpose, but in this case the resultant 
whole controlled system is not a minimum-noise amplifier. 
Instead, this paper proposes the following feedback connection as shown 
in Fig.~2: 
\begin{equation}
\label{feedback connection}
      \tilde{b}_2 =b_3,~~~
      b_2=\tilde{b}_4, 
\end{equation}
which is, of course, equivalent to $\tilde{b}_2^\dagger=b_3^\dagger$ and 
$b_2^\dagger=\tilde{b}_4^\dagger$. 
Note that in Fig.~2 practical unwanted noises $d_3, \ldots, d_6$ are 
illustrated, but these modes are ignored for the moment. 
From Eqs.~\eqref{system}, \eqref{controller} and \eqref{feedback connection}, 
the whole controlled system, with inputs $b_1, b_4^\dagger$ and outputs 
$\tilde{b}_1, \tilde{b}_3^\dagger$, has the following input-output relation 
in the Laplace domain: 
\[
      \left[ \begin{array}{c}
                \tilde{b}_1(s) \\ 
                \tilde{b}_3^\dagger(s) \\
            \end{array} \right]
          = \left[ \begin{array}{cc}
               G_{11}^{({\rm fb})}(s) & G_{12}^{({\rm fb})}(s) \\
               G_{21}^{({\rm fb})}(s) & G_{22}^{({\rm fb})}(s) \\
             \end{array} \right]
           \left[ \begin{array}{c}
                 b_1(s) \\ 
                 b_4^\dagger(s) \\
             \end{array} \right], 
\]
where 
\begin{eqnarray*}
& & \hspace*{-1em}
      G_{11}^{({\rm fb})} 
          = [G_{11} - K_{21}(G_{11}G_{22}-G_{12}G_{21})]/(1-K_{21}G_{22}),
\nonumber \\ & & \hspace*{-1em}
      G_{12}^{({\rm fb})} = (G_{12} K_{22})/(1-K_{21}G_{22}), 
\nonumber \\ & & \hspace*{-1em}
      G_{21}^{({\rm fb})} = (G_{21} K_{11})/(1-K_{21}G_{22}), 
\nonumber \\ & & \hspace*{-1em}
      G_{22}^{({\rm fb})} 
          = [K_{12} + G_{22}(K_{11}K_{22}-K_{12}K_{21}) ]/(1-K_{21}G_{22}). 
\end{eqnarray*}
The matrix entries satisfy the condition corresponding to Eq.~\eqref{G conditions}, 
i.e., $|G_{11}^{({\rm fb})}(i\omega)|^2 - |G_{12}^{({\rm fb})}(i\omega)|^2 =1$ 
$\forall \omega$, etc. 
Finally, as remarked in Sec.~II, for the whole controlled system to be stable, 
the controller should be carefully designed so that all poles of $G_{ij}^{({\rm fb})}(s)$ 
must lie in the left-hand complex plane, as demonstrated in Sec.~V.


\subsection{Robust amplification via feedback}

We now focus on the output of the controlled system in the Fourier domain, i.e., 
\[
     \tilde{b}_1(i\omega) = G_{11}^{({\rm fb})}(i\omega)b_1(i\omega)
               + G_{12}^{({\rm fb})}(i\omega)b_4^\dagger(i\omega),
\]
and the amplification gain $|G_{11}^{({\rm fb})}(i\omega)|$ especially when 
the original gain $|G_{11}(i\omega)|$ is large. 
Note that $G_{11}^{({\rm fb})}$ looks somewhat different from the classical 
counterpart \eqref{classical FB}; hence, it is not immediate to see if 
$|G_{11}^{({\rm fb})}(i\omega)|$ can be approximated by a function of 
only the controller. 
Nonetheless, the analogous result to the classical case indeed holds as shown 
below.

For the proof we use Eq.~\eqref{G conditions} (below, we omit the variable 
$i\omega$). 
First, from $|G_{21}||G_{11}|=|G_{22}||G_{12}|$ together with the other 
two equations, we  have 
$|G_{11}|=|G_{22}|$ and $|G_{12}|=|G_{21}|$. 
Also $G_{11}G_{22}-G_{12}G_{21} = G_{22}/G_{11}^*$ holds. 
Here, in the limit $|G_{11}|\rightarrow \infty$, it follows that 
\[
       \frac{|G_{11}G_{22}-G_{12}G_{21}|}{|G_{11}|}
        = \frac{|G_{22}|}{|G_{11}|^2}
        = \frac{1}{|G_{11}|} \rightarrow 0.
\]
This implies that $(G_{11}G_{22}-G_{12}G_{21})/|G_{11}|$ converges to zero 
in this limit. 
As a consequence, we have 
\begin{eqnarray*}
& & \hspace*{-1em}
      |G_{11}^{({\rm fb})}| 
          = \Big|\frac{G_{11}/|G_{11}| - K_{21}(G_{11}G_{22}-G_{12}G_{21})/|G_{11}|}
                      {1/|G_{11}|-K_{21}G_{22}/|G_{11}|} \Big|
\nonumber \\ & & \hspace*{2em}
      \rightarrow 
       \Big|\frac{G_{11}/|G_{11}|}{-K_{21}G_{22}/|G_{11}|} \Big|
       = \frac{1}{|K_{21}|}.
\end{eqnarray*}
Hence, in the frequency range where the plant has a large gain 
$|G_{11}(i\omega)|\gg 1$, the whole controlled system amplifies the input 
$b_1(i\omega)$ with gain 
$|G_{11}^{({\rm fb})}(i\omega)| \approx 1/|K_{21}(i\omega)|>1$. 
Therefore we obtain the desirable quantum robust amplification method via 
feedback; 
that is, thanks to the fact that the passive controller is much more robust 
compared to the original amplifier, even if $G_{11}$ changes while maintaining 
a large value, the whole controlled system carries out robust amplification 
with stable gain $1/|K_{21}|$.


\subsection{Feedback gain synthesis}

Here we conduct a quantitative analysis on the robustness property, which 
provides a guideline for synthesizing the feedback gain $K$. 
To see the idea clearly, let us again consider the classical case \eqref{classical FB}. 
Let $\Delta G$ be the fluctuation that occurs in the plant $G$; 
then the fluctuation that occurs in the whole controlled system $G^{\rm (fb)}$ 
is calculated as 
\[
     \Delta G^{\rm (fb)} 
       = \frac{G+\Delta G}{1+(G+\Delta G)K} - \frac{G}{1+GK}
       \approx \frac{\Delta G}{(1+GK)^2}, 
\]
which as a result leads to 
\begin{equation}
\label{classical sensitivity reduction}
     \frac{\Delta G^{\rm (fb)}}{G^{\rm (fb)}} 
      = \frac{1}{1+GK} \cdot \frac{\Delta G}{G}. 
\end{equation}
Hence, the gain sensitivity to the unwanted fluctuation can be reduced by 
the factor $1/|1+GK|$ by feedback. 
Equation \eqref{classical sensitivity reduction} suggests to us not to design 
$G$ and $K$ separately; 
rather, what determines the performance of the controlled amplifier 
is the {\it loop gain} $GK$. 
Actually, while the controlled amplification gain, $G^{\rm (fb)}\approx 1/K$, 
can be made bigger by taking a smaller value of $K$, we should not design a 
too small $K$ such that $GK\approx 0$; 
in this case, Eq.~\eqref{classical sensitivity reduction} yields 
$\Delta G^{\rm (fb)}/G^{\rm (fb)} = \Delta G/G$, and thus there is no 
improvement in the sensitivity. 
The so-called {\it Bode plot} developed by Bode (e.g., see Ref. \cite{Bode book}) 
is a powerful graphical method for synthesizing $K$ as well as $G$, and it is 
now the standard tool for general feedback circuit design.

Now let us try to establish the quantum version of the above discussion. 
Note in advance that a straightforward calculation, like 
Eq.~\eqref{classical sensitivity reduction}, cannot be carried out in the quantum 
case, but nonetheless a similar useful equation for determining the controller 
parameter $K_{21}$ is shown. 
First, due to $|G_{11}|=|G_{22}|$ and 
$G_{11}G_{22}-G_{12}G_{21} = G_{22}/G_{11}^*$, we find 
\[
       G_{11}^{\rm (fb)}
        = \frac{|G_{11}|^2 - K_{21}G_{22}}{G_{11}^*(1-K_{21}G_{22})}
        = \frac{G_{22}}{G_{11}^*} \cdot \frac{G_{22}^* - K_{21}}{1-K_{21}G_{22}}, 
\]
which thus leads to 
\[
       | G_{11}^{\rm (fb)} |
        = \frac{| G_{22}^* - K_{21} |}{ | 1-K_{21}G_{22} |}. 
\]
The fluctuation of the controlled gain is given by 
\[
     \Delta | G_{11}^{\rm (fb)} |
        = \frac{| (G_{22} + \Delta G_{22})^* - K_{21} |}
                    { | 1-K_{21}(G_{22}+\Delta G_{22}) |}
          - \frac{| G_{22}^* - K_{21} |}{ | 1-K_{21}G_{22} |}. 
\]
Then, from the general relation 
$|x+\epsilon|\approx|x|+(x\epsilon^*+x^*\epsilon)/2|x|$ with 
$x, \epsilon\in {\mathbb C}$ and $|\epsilon|\ll 1$, the normalized fluctuation 
of the amplification gain of the controlled system can be explicitly calculated as 
\begin{equation}
\label{explicit robust}
     \frac{\Delta|G_{11}^{\rm (fb)}|}{|G_{11}^{\rm (fb)}|}
      = \frac{1-|K_{21}|^2}{|G_{22}^*-K_{21}|^2} \cdot 
           {\rm Re}\Big( 
             \frac{G_{22}^*-K_{21}}{1-K_{21}G_{22}}\Delta G_{22} \Big). 
\end{equation}
Next, noting that $|G_{22}|=|G_{11}|\gg 1$ and thus 
$|K_{21}|\ll |G_{22}|$, we find from Eq.~\eqref{explicit robust} that 
\[
     \Big| \frac{\Delta|G_{11}^{\rm (fb)}|}{|G_{11}^{\rm (fb)}|} \Big|
      \approx \frac{1}{|G_{22}|^2} \cdot 
           \Big| {\rm Re}\Big( 
             \frac{G_{22}^*}{1-K_{21}G_{22}}\Delta G_{22} \Big) \Big|.
\]
Thus, from the general relation $|{\rm Re}(xy)|\leq |x| |y|$ 
with $x, y\in {\mathbb C}$, it follows that 
\begin{equation}
\label{robust ineq}
      \Big| \frac{\Delta|G_{11}^{\rm (fb)}|}{|G_{11}^{\rm (fb)}|} \Big|
      \leq \frac{1}{|1-K_{21}G_{22}|} \cdot \frac{|\Delta G_{22}|}{|G_{22}|}. 
\end{equation}

Equation \eqref{robust ineq} has a similar form to 
Eq.~\eqref{classical sensitivity reduction} and indeed provides us a guideline 
for feedback design. 
That is, as in the classical case, the balance of $K_{21}$ and $G_{22}$ determines 
the ability of the controlled system to suppress the fluctuation 
(note that $|\Delta G_{22}|\geq \Delta |G_{22}|=\Delta |G_{11}|$). 
In particular, we now deduce a similar conclusion as in the classical case; 
if we choose a too small $K_{21}$ in addition to $|G_{11}|\gg 1$ such that the 
loop gain $K_{21}G_{22}$ is almost zero, then substituting the relation 
\[
      \Delta |G_{11}|=\Delta |G_{22}|
         ={\rm Re}(G_{22}^*\Delta G_{22})/|G_{22}|
\]
into Eq.~\eqref{explicit robust} we obtain 
$\Delta|G_{11}^{\rm (fb)}|/|G_{11}^{\rm (fb)}|=\Delta|G_{11}|/|G_{11}|$. 
That is, in this case the fluctuation is not at all suppressed via feedback. 
To design an appropriate controller gain $K_{21}$, the Bode plot of the 
loop gain $K_{21}(i\omega)G_{22}(i\omega)$ is useful.


\section{Quantum noise limit}

Let us define the noise magnitude of $b$ by 
\[
     \mean{|\Delta b|^2}
            :=\frac{1}{2}
                 \mean{\Delta b\Delta b^\dagger + \Delta b^\dagger \Delta b},~~~
       \Delta b=b-\mean{b}. 
\]
Then, through the ideal amplification process 
$\tilde{b}_1 = g_1 b_1 + g_2 b_2^\dagger$, the noise magnitude must be also 
amplified as 
$\mean{|\Delta \tilde{b}_1|^2} = |g_1|^2 \mean{|\Delta b_1|^2} + |g_2|^2/2$, 
where $b_2$ is assumed to be in the vacuum. 
This implies the degradation of the signal-to-noise ratio: 
\begin{eqnarray*}
& & \hspace*{-1em}
    \widetilde{{\rm (S/N)}}=
    \frac{|\mean{\tilde{b}_1}|^2}{\mean{|\Delta \tilde{b}_1|^2}}
          = \frac{|\mean{b_1}|^2}
                      {\mean{|\Delta b_1|^2} + {\cal A}}
          < \frac{|\mean{b_1}|^2}{\mean{|\Delta b_1|^2}} 
          = {\rm (S/N)}. 
\end{eqnarray*}
Hence, the {\it added noise} 
\[
       {\cal A} := \frac{|g_2|^2}{2|g_1|^2}
        = \frac{|g_1|^2-1}{2|g_1|^2}
\]
quantifies the fidelity of the amplification process \cite{Haus 62,Caves 82}. 
In particular, in the large amplification limit $|g_1|\rightarrow \infty$ 
we find ${\cal A}\rightarrow 1/2$, which is called the {\it quantum noise limit}.

Up to now, the ideal setup is assumed, and the controlled system is driven by 
only the signal $b_1$ and the auxiliary input $b_4$, implying that it actually 
reaches the quantum noise limit in the large amplification limit. 
Hence, here we consider the following general case where some excess noise 
exists, as illustrated in Fig.~2; 
the plant is subjected to an unwanted noise $d_3$ that enters into 
the system in the form $\tilde{b}_1=G_{11}b_1+G_{12}b_2^\dagger+G_{13}d_3$; 
the controller is also affected by a noise $d_4$; 
furthermore, the feedback transmission lines are lossy, which is modeled by 
inserting fictitious beam splitters with additional inputs $d_5$ and $d_6$. 
Note that $d_3, \ldots, d_6$ are all annihilation modes. 
Then the output of the whole controlled system has the form 
\[
     \tilde{b}_1 = G_{11}^{\rm (fb)}b_1 + G_{12}^{\rm (fb)}b_2^\dagger 
                + G_{13}^{\rm (fb)}d_3 + G_{14}^{\rm (fb)}d_4^\dagger 
                + G_{15}^{\rm (fb)}d_5^\dagger + G_{16}^{\rm (fb)}d_6^\dagger. 
\]
Then, if the excess noises are all vacuum, the added noise in the 
feedback-controlled amplification process, denoted by ${\cal A}^{\rm (fb)}$, 
satisfies 
\begin{equation}
\label{imp added noise}
      \lim_{|G_{11}^{\rm (fb)}|\rightarrow \infty}{\cal A}^{\rm (fb)} 
        = \lim_{|G_{11}|\rightarrow \infty}{\cal A}^{\rm (o)}
        = \frac{1}{2} + \frac{|G_{13}|^2}{|G_{11}|^2},
\end{equation}
where ${\cal A}^{\rm (o)}$ is the added noise of the plant. 
The proof of Eq.~\eqref{imp added noise}, including the detailed forms of 
$G_k^{\rm (fb)}$, is given in Appendix~A. 
This is a very useful result for the following reasons. 
First, in the large amplification limit the two added noises ${\cal A}^{\rm (fb)}$ 
and ${\cal A}^{\rm (o)}$ are equal; 
as a consequence the second term in the right-hand side of 
Eq.~\eqref{imp added noise} is a function of only the plant and 
cannot be further altered by feedback control. 
Hence, we have the following no-go theorem: 
If the original amplifier does not reach the quantum noise limit 
(i.e., $\lim_{|G_{11}|\rightarrow \infty}|G_{13}|/|G_{11}|> 0$), 
the controller can never remove this excess noise. 
On the other hand, notably, the imperfections contained in the controller 
and the feedback transmission lines do not appear in Eq.~\eqref{imp added noise}. 
This means that a very accurate fabrication of the feedback controller is 
not necessarily required, which is a desirable fact from a practical viewpoint. 
Thus, if the original amplifier operates with the minimum added noise, the 
controlled system reaches the quantum noise limit as well even if some 
imperfections are present in the feedback loop.


\section{Application to optical broadband amplification}

In any practical situation, it is important to carefully engineer an amplifier 
so that it has a proper frequency bandwidth in which nearly constant 
amplification gain is realized. 
On the other hand, it is known in both the classical and quantum cases that, 
particularly for an amplifier with a single pole, the effective bandwidth becomes 
smaller if the amplification gain is taken to be bigger. 
That is, there is a {\it gain-bandwidth constraint}. 
However, this constraint is not necessarily applied to a more complex amplifier 
with multiple poles. 
In fact, recently in Ref. \cite{Clerk 14}, the authors propose a hybrid amplifier 
composed of two cavity modes and an additional opto-mechanical mode that 
circumvents the gain-bandwidth constraint. 
In this section, we study another system that is also free from this constraint, 
that yet does not need an additional degree of freedom. 
Then, the effectiveness of feedback is discussed, demonstrating its ability to 
make the system robust.


\subsection{NDPA with special detuning}

\begin{figure}[t]
\centering 
\includegraphics[width=8.8cm]{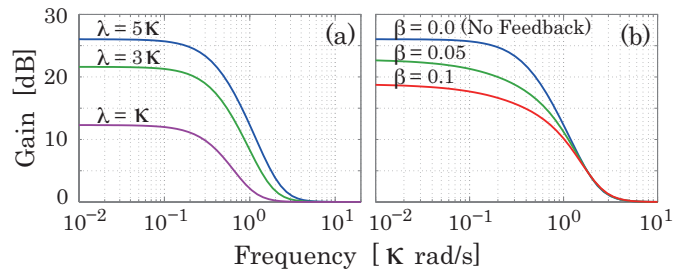}
\caption{
(a) Gain profile of the specially detuned NDPA without feedback. 
(b) Gain profile of the feedback-controlled system with parameters 
$\lambda=5\kappa$ and various $\beta$. 
}
\label{examples}
\end{figure}

The plant system is the NDPA with dynamics \eqref{NDPA dynamics 1}, 
\eqref{NDPA dynamics 2} and output \eqref{NDPA output}. 
Here we consider the ideal case where the unwanted noises $d_3, \ldots, d_6$ 
shown in Fig.~2 are not present. 
Without any invention, this system is subject to a gain-bandwidth constraint 
\cite{Clerk 10}, but now let us take the specific detuning as 
$\Delta_1=\Delta_2=\lambda$. 
The transfer function matrix of this system is then given by 
\[
       G(s) = 
       \frac{1}{(s+\kappa/2)^2}
       \left[ \begin{array}{cc}
               s^2-\kappa^2/4 + i\kappa \lambda & -\kappa \lambda \\
               -\kappa \lambda & s^2-\kappa^2/4 - i\kappa \lambda \\
             \end{array} \right].
\]
The maximum gain is $|G_{11}(0)|=\sqrt{1+16\lambda^2/\kappa^2}$, 
which becomes larger by increasing $\lambda$. 
Remarkably, this amplification can be carried out without sacrificing the 
bandwidth. 
Figure~3 (a) shows the three cases corresponding to 
$\lambda=\kappa, 3\kappa, 5\kappa$, all of which have the same 
effective bandwidth $\sim \kappa/10$. 
A clear advantage of this system is in its implementability; that is, it is 
composed of only optical devices, and there is no need to prepare an auxiliary 
system such as an opto-mechanical oscillator. 
Also, note that the system is always stable (the pole of the transfer function 
matrix is $-\kappa/2$); that is, in a proper parameter regime such that 
the linearized model given by Eqs.~\eqref{NDPA dynamics 1} and 
\eqref{NDPA dynamics 2} is valid 
\footnote{
In order for the rotating-wave approximation to hold, we require that the 
detuning $\Delta_1=\Delta_2=\lambda$ must be much smaller than the 
optical frequencies $(\omega_0, \omega_1, \omega_2)$ (see Sec.~II). 
Also, in a strong coupling regime such that $\lambda\gg \kappa$ holds, 
the third-order nonlinearity contained in the Hamiltonian becomes effective, 
and the linearized model \eqref{NDPA dynamics 1} and \eqref{NDPA dynamics 2} 
is then not valid anymore. 
Taking into account these facts, in the simulation, we take a relatively small 
value of $\lambda$, i.e., $\lambda=\kappa, 3\kappa, 5\kappa$, which is 
typically of the order of megahertz. 
}, 
there is no clear upper bound on $\lambda$, in contrast to the standard NDPA 
which imposes $|\lambda|<\kappa/2$.


\subsection{The feedback effect}

\begin{figure}[t]
\centering 
\includegraphics[width=6.8cm]{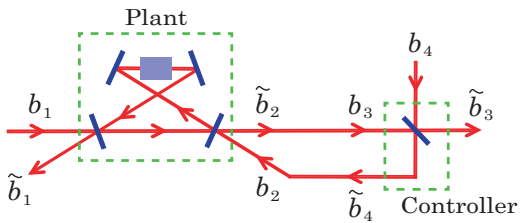}
\caption{
The NDPA (plant) and its coherent feedback; 
by simply feeding the auxiliary output $\tilde{b}_2$ back to the auxiliary 
input $b_2$ through just a beam splitter, the robustness property is drastically 
improved, as discussed in the main text. 
}
\label{examples}
\end{figure}

Next, let us consider the feedback control of this amplifier, again in the ideal setup. 
Here, as shown in Fig.~4, a beam splitter with transmissivity $\alpha$ and 
reflectivity $\beta$ is taken as a controller. 
This device has no internal dynamics, and its transfer function matrix is constant: 
\[
       K(s) = 
       \left[ \begin{array}{cc}
               \alpha & \beta \\
               \beta & -\alpha \\
             \end{array} \right],~~~
        \alpha, \beta \in{\mathbb R}. 
\]
Thus, $K_{21}=\beta$ represents the attenuation level. 
The amplification gain of the whole controlled system is then 
\[
      G_{11}^{({\rm fb})}(s)
       = \frac
            {(1-\beta)s^2 + \beta\kappa s - (1+\beta)\kappa^2/4 + i \kappa \lambda}
            {(1-\beta)s^2 + \kappa s + (1+\beta)\kappa^2/4 + i\beta \kappa \lambda}.
\]
This expression shows that, as expected from the general theory discussed in 
Sec.~III B, in the limit 
$|G_{11}(i\omega)|\rightarrow \infty$ (i.e., $\lambda\rightarrow \infty$) 
the gain becomes $|G_{11}^{({\rm fb})}(i\omega)|\rightarrow 1/\beta$ in 
a certain frequency bandwidth. 
To determine the attenuation level $\beta$, we need the stability condition; 
it is shown in Appendix~B that, for all the poles of $G_{11}^{({\rm fb})}(s)$ 
to lie in the left-hand complex plane, the parameters must satisfy 
\begin{equation}
\label{stability condition}
      |\lambda| < \frac{\kappa}{2}\sqrt{\frac{1+\beta}{\beta^2(1-\beta)}}. 
\end{equation}
This yields $|\beta| < \kappa/2|\lambda|$ when $\beta^2\ll 1$; 
hence, let us here choose $\lambda=5\kappa$, leading to $|\beta|< 0.1$. 
Figure 3~(b) shows the gain $|G_{11}^{({\rm fb})}(i\omega)|$ for the two 
cases $\beta=0.1$ and $\beta=0.05$ together with the plot without feedback 
(i.e., $\beta=0$). 
We then observe that the gain of the controlled amplifier becomes smaller 
than that without feedback; in exchange for this reduced gain, the controlled 
amplifier obtains a great robustness property against the parameter fluctuation 
as demonstrated later. 
Note that a larger value of $\beta$ (thus smaller amplification gain) induces 
a wider frequency bandwidth; 
hence, the controlled NDPA has the gain-bandwidth constraint. 
But the point here is rather that the gain and bandwidth can be easily tuned 
by just changing the reflectivity of the beam splitter. 
That is, an easily adjustable amplification can be realized, and this is also 
a clear advantage of feedback.


\subsection{Robustness property} 

\begin{figure}[t]
\centering 
\includegraphics[width=8.7cm]{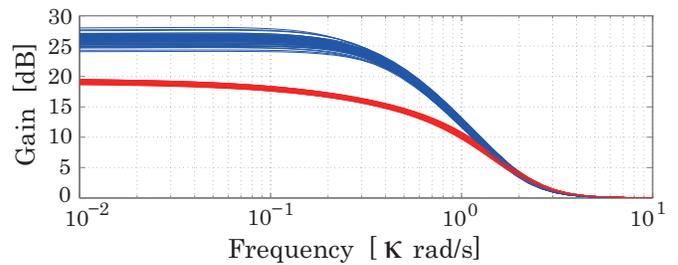}
\caption{
The upper blue lines represent the gain profile of the specially detuned NDPA 
without feedback, $|G_{11}(i\omega)|$, while the lower red lines correspond 
to the controlled case $|G_{11}^{\rm (fb)}(i\omega)|$ with $\beta=0.1$. 
In both cases, $\lambda_0=5\kappa$. 
}
\label{examples}
\end{figure}

As repeatedly emphasized in this paper, the main strength of feedback 
is that the controlled system possesses a robustness property. 
To see this, let us consider an imperfect case as follows. 
First, the device parameters are fragile; the coupling strength $\lambda$ 
fluctuates in such a way that $\lambda=(1+0.1\epsilon_0)\lambda_0$, 
where $\lambda_0$ is the nominal value; 
similarly, the detunings $\Delta_1$ and $\Delta_2$ can be slightly deviated 
from $\lambda$, which is modeled by 
$\Delta_1=(1+0.001\epsilon_1)\lambda$ and 
$\Delta_2=(1+0.001\epsilon_2)\lambda$. 
Here $(\epsilon_0, \epsilon_1, \epsilon_2)$ are independent random variables 
subjected to the uniform distribution in $[-1, 1]$. 
In addition to this fragility, we assume that the signal mode $a_1$ is subjected 
to optical loss, which is modeled by adding the extra term 
$-\gamma a_1/2-\sqrt{\gamma}d_3$ to the right-hand side of 
Eq.~\eqref{NDPA dynamics 1}, with $\gamma$ the magnitude of the loss and 
$d_3$ the unwanted vacuum noise. 
The feedback transmission lines are also lossy, which is modeled by 
Eq.~\eqref{imperfect FB connection}.

The blue lines in Fig.~5 are 50 sample values of the autonomous gain 
$|G_{11}(i\omega)|$ in the case $\lambda_0=5\kappa$ and 
$\alpha_1=\alpha_2=0.99$. 
That is, in fact, due to the parameter fluctuation described above, the 
amplifier becomes fragile and the amplification gain significantly varies. 
Nonetheless, this fluctuation can be suppressed by feedback; 
the red lines in Fig.~5 are 50 sample values of the controlled gain 
$|G_{11}^{\rm (fb)}(i\omega)|$ with attenuation level $\beta=0.1$, whose 
fluctuation is indeed much smaller than that of $|G_{11}(i\omega)|$ 
\footnote{
This result can be predicted by Eq.~\eqref{robust ineq}, which guarantees 
that the fluctuation at the center frequency $\omega=0$ can be suppressed 
at least by a factor of $1/|1-K_{21}(0)G_{22}(0)|=0.44$. 
But this means that Eq.~\eqref{robust ineq} is conservative, since Fig.~5 
shows that the fluctuation is suppressed about by a factor of 0.2.}. 
(Note that, because the fluctuation of $|G_{11}^{\rm (fb)}(i\omega)|$ is 
very small, the set of sample values looks like a thick line.) 
That is, the controlled system is certainly robust against the realistic 
fluctuation of the device parameters.


\subsection{Added noise}

\begin{figure}[t]
\centering 
\includegraphics[width=8.8cm]{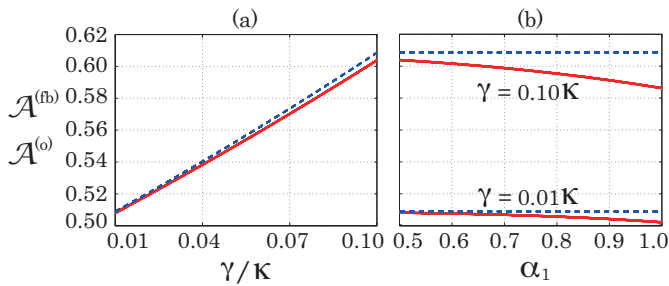}
\caption{
The added noise of the controlled amplifier with attenuation level $\beta=0.1$ 
and that of the non-controlled one (i.e., $\beta=0$) versus 
(a) $\gamma/\kappa$ with fixed $\alpha_1=\alpha_2=0.5$, and 
(b) $\alpha_1=\alpha_2$ with fixed $\gamma/\kappa$. 
In both figures, the red solid lines represent ${\cal A}^{\rm (fb)}$, while the 
blue dotted lines are ${\cal A}^{\rm (o)}$, at $\omega=0$. 
}
\label{examples}
\end{figure}

Finally, let us investigate how much the excess noise is added to the output of 
the controlled or noncontrolled specially detuned NDPA. 
Again we set $\lambda_0=5\kappa$, and the feedback control is conducted 
with attenuation level $\beta=0.1$. 
Also, the same imperfections considered in the previous subsection are assumed; 
that is, the system suffers from the signal loss (represented with $\gamma$) 
and the probabilistic fluctuation of the parameters 
($\lambda, \Delta_1, \Delta_2$); 
furthermore, the feedback control is implemented with the lossy transmission lines 
(represented with $\alpha_1$ and $\alpha_2$).

With this setup Fig.~6 is obtained, where the red solid lines are sample values 
of the added noise at the center frequency $\omega=0$ for the controlled 
system, ${\cal A}^{\rm (fb)}$ given by Eq.~\eqref{realistic A}, while the blue 
dotted lines represent those of the non-controlled system, ${\cal A}^{\rm (o)}$ 
given by Eq.~\eqref{realistic Ao}. 
(In the figure it appears that six thick lines are plotted, but each is the set of 
50 sample values.) 
Figure~6~(a) shows ${\cal A}^{\rm (fb)}$ and ${\cal A}^{\rm (o)}$ versus the 
signal loss rate $\gamma/\kappa$, where for the controlled system we fix 
$\alpha_1=\alpha_2=0.5$ (that is, the feedback transmission lines are 
very lossy). 
Also, Fig.~6~(b) shows the added noise as a function of $\alpha_1=\alpha_2$, 
with fixed signal loss $\gamma/\kappa$.

The first crucial point is that, in both figures (a) and (b), ${\cal A}^{\rm (fb)}$ 
and ${\cal A}^{\rm (o)}$ are close to each other. 
This is the fact that can be expected from Eq.~\eqref{imp added noise}, which 
states that ${\cal A}^{\rm (fb)}$ and ${\cal A}^{\rm (o)}$ coincide in the large 
amplification limit. 
It is also notable that, for all sample values, ${\cal A}^{\rm (fb)}$ is smaller 
than ${\cal A}^{\rm (o)}$ 
\footnote{
This fact can be roughly understood in the following way. 
Let us focus on Eq.~\eqref{11_13 ratio}, particularly $G_{11}G_{22}-G_{12}G_{21}$ 
in the denominator and $G_{13}G_{22}-G_{12}G_{23}$ in the numerator. 
Then $|G_{11}G_{22}-G_{12}G_{21}|^2$ is, from Eq.~\eqref{proof eq 1}, 
upper bounded by $|G_{11}|^2+|G_{22}|^2$, while, from Eq.~\eqref{proof eq 2}, 
$|G_{13}G_{22}-G_{12}G_{23}|^2$ is upper bounded by $|G_{22}|^2$. 
Hence, if both of these upper bounds are reached, together with the relation 
$|G_{13}|<|G_{11}|$ observed in the figure, Eq.~\eqref{11_13 ratio} yields 
$|G_{13}^{\rm (fb)}|^2/|G_{11}^{\rm (fb)}|^2 < |G_{13}|^2/|G_{11}|^2$, 
which means ${\cal A}^{\rm (fb)}<{\cal A}^{\rm (o)}$ in the large 
amplification limit. 
}
; in other words, the feedback controller reduces the added noise, although 
in the large amplification limit this effect becomes negligible as proven in 
Eq.~\eqref{imp added noise}. 
Another important feature is that, as seen in Fig.~6~(a), the signal loss 
$\gamma$ is the dominant factor increasing the added noise, and the 
feedback loss $1-\alpha_1 (= 1-\alpha_2)$ does not have a large impact 
on it, as seen in Fig.~6~(b). 
As consequence, when $\gamma$ is small, the controlled amplifier can 
perform amplification nearly at the quantum noise limit $1/2$, with almost 
no dependence on the feedback loss; 
this fact is also consistent with Eq.~\eqref{imp added noise}.

In summary, the specially detuned NDPA with feedback control functions 
as a robust, near-minimum-noise (if $\gamma\ll \kappa$), and broadband 
amplifier.


\section{Conclusion}

The presented feedback control theory resolves the critical fragility issue in 
phase-preserving linear quantum amplifiers. 
The theory is general and thus applicable to many different physical setups, 
such as optics, opto-mechanics, superconducting circuits, and their hybridization. 
Moreover, the feedback scheme is simple and easy to implement, as 
demonstrated in Sec. V. 
Note also that the case of {\it phase-conjugating amplification} 
\cite{Cerf 01,Leuchs 2007} can be discussed in a similar way; see Appendix~C.

In a practical setting, the controller synthesis problem becomes complicated, 
implying the need to develop a more sophisticated quantum feedback amplification 
theory, which indeed was established in the classical case 
\cite{Mancini,Bode book,Astrom Murray}. 
The combination of those classical approaches with the quantum control theory 
\cite{Doherty 2000,WisemanBook,KurtBook} should advance this research 
direction. 
Another interesting future work is to study genuine quantum-mechanical 
settings, e.g., probabilistic amplification 
\cite{Caves 13,Ralph 09,Pryde 10,Ralph 14}. 
Finally, note that feedback control is used in order to reach the quantum noise 
limit, in a different amplification scheme (the so-called op-amp mode) 
\cite{Clerk 10,Courty 99}; 
connection to these works is also to be investigated.

This work was supported in part by JSPS Grant-in-Aid No. 15K06151.


\appendix


\section{Proof of Eq.~(12)}

First, we again describe the setup of the imperfect system depicted in Fig.~2, 
in a more detailed way. 
The plant system is subjected to an unwanted noise $d_3$, such that the 
input-output relationship is given by 
\begin{equation}
\label{imperfect system}
      \left[ \begin{array}{c}
                \tilde{b}_1(s) \\ 
                \tilde{b}_2^\dagger(s) \\
            \end{array} \right]
         = \left[ \begin{array}{ccc}
             G_{11}(s) & G_{12}(s) & G_{13}(s) \\
             G_{21}(s) & G_{22}(s) & G_{23}(s) \\
            \end{array} \right]
        \left[ \begin{array}{c}
                     b_1(s) \\ 
                     b_2^\dagger(s) \\
                     d_3(s) \\ 
                 \end{array} \right]. 
\end{equation}
The transfer function matrix in this case satisfies 
\begin{align}
& \hspace*{0em}
    |G_{11}|^2 - |G_{12}|^2 + |G_{13}|^2 =
    |G_{22}|^2 - |G_{21}|^2 - |G_{23}|^2 = 1,
\nonumber \\ 
& \hspace*{0em}
    G_{21}G_{11}^* - G_{22}G_{12}^* + G_{23}G_{13}^* = 0, 
\label{imperfect G conditions}
\end{align}
for all $s=i\omega$. 
Thus, the added noise of the plant system is given by 
\begin{eqnarray}
\label{realistic Ao}
& & \hspace*{-0.9em}
    {\cal A}^{\rm (o)} 
          = \frac{|G_{12}|^2 + |G_{13}|^2}{2|G_{11}|^2}
          = \frac{|G_{11}|^2 + 2|G_{13}|^2 -1 }{2|G_{11}|^2}
\nonumber \\ & & \hspace*{1.2em}
     = \frac{1}{2} - \frac{1}{2|G_{11}|^2}
         + \frac{|G_{13}|^2}{|G_{11}|^2}. 
\end{eqnarray}
This leads to 
\begin{equation}
\label{realistic Ao limit}
      \lim_{|G_{11}|\rightarrow\infty}
             {\cal A}^{\rm (o)} = \frac{1}{2} + \frac{|G_{13}|^2}{|G_{11}|^2}, 
\end{equation}
where the second term is assumed to exist. 
Note that in a more general setup some creation noise modes (e.g. $d_3^\dagger$) 
can be contained in Eq.~\eqref{imperfect system}, but this modification does 
not change the conclusion. 
The controller also contains an unwanted noise field $d_4$ in the following form:
\begin{equation}
\label{imperfect controller}
      \left[ \begin{array}{c}
                \tilde{b}_3^\dagger(s) \\ 
                \tilde{b}_4^\dagger(s) \\
            \end{array} \right]
       = \left[ \begin{array}{ccc}
                     K_{11}(s) & K_{12}(s) & K_{13}(s) \\
                     K_{21}(s) & K_{22}(s) & K_{23}(s) \\
                   \end{array} \right]
       \left[ \begin{array}{c}
                     b_3^\dagger(s) \\ 
                     b_4^\dagger(s) \\
                     d_4^\dagger(s) \\
                   \end{array} \right].
\end{equation}
The point here is that only the creation modes appear in 
Eq.~\eqref{imperfect controller}, unlike Eq.~\eqref{imperfect system} that 
involves both creation and annihilation modes; this is indeed due to the 
passivity property of the controller. 
Finally, the transmission lines (optical fields) for feedback are assumed to 
be lossy. 
This setting is modeled by inserting two fictitious beam splitters; 
the beam splitter in the output field $\tilde{b}_2$ has transmissivity 
$\alpha_1$ and reflectivity $\delta_1$, and also the beam splitter in 
the output field $\tilde{b}_4$ has transmissivity $\alpha_2$ and reflectivity 
$\delta_2$ (we assume $\alpha_i, \delta_i\in{\mathbb R}$ without loss of 
generality). 
Then, the coherent feedback connection is represented by the following 
relations: 
\begin{equation}
\label{imperfect FB connection}
     b_3^\dagger=\alpha_1 \tilde{b}_2^\dagger + \delta_1 d_5^\dagger,~~~
     b_2^\dagger=\alpha_2 \tilde{b}_4^\dagger + \delta_2 d_6^\dagger. 
\end{equation}
Combining Eqs.~\eqref{imperfect system}, \eqref{imperfect controller}, 
and \eqref{imperfect FB connection}, we end up with 
\[
     \tilde{b}_1 = G_{11}^{\rm (fb)}b_1 + G_{12}^{\rm (fb)}b_4^\dagger 
                + G_{13}^{\rm (fb)}d_3 + G_{14}^{\rm (fb)}d_4^\dagger 
                + G_{15}^{\rm (fb)}d_5^\dagger + G_{16}^{\rm (fb)}d_6^\dagger, 
\]
where 
\begin{eqnarray*}
& & \hspace*{-0.6em}
      G_{11}^{\rm (fb)} 
       = [ G_{11}-\alpha_1\alpha_2 K_{21} (G_{11}G_{22}-G_{12}G_{21}) ]/ {\bf G}, 
\nonumber \\ & & \hspace*{-0.6em}
      G_{13}^{\rm (fb)} 
       = [ G_{13}-\alpha_1\alpha_2 K_{21} (G_{13}G_{22}-G_{12}G_{23}) ]/ {\bf G},
\nonumber \\ & & \hspace*{-0.6em}
      G_{12}^{\rm (fb)} 
       = \alpha_2 G_{12} K_{22} / {\bf G},~~~
      G_{14}^{\rm (fb)} 
       = \alpha_2 G_{12} K_{23} / {\bf G},
\nonumber \\ & & \hspace*{-0.6em}
      G_{15}^{\rm (fb)} 
       = \alpha_2 \delta_1 G_{12} K_{21} / {\bf G},~~~
      G_{16}^{\rm (fb)} 
       = \delta_2 G_{12} / {\bf G}, 
\end{eqnarray*}
and ${\bf G}=1-\alpha_1\alpha_2 K_{21}G_{22}$. 
Note that the transfer functions satisfy at $s=i\omega$ 
\begin{eqnarray}
& & \hspace*{-0.6em}
\label{imperfect constraint}
     |G_{11}^{\rm (fb)}|^2 - |G_{12}^{\rm (fb)}|^2 + |G_{13}^{\rm (fb)}|^2 
\nonumber \\ & & \hspace*{2em}
      \mbox{}
       -|G_{14}^{\rm (fb)}|^2 - |G_{15}^{\rm (fb)}|^2 - |G_{16}^{\rm (fb)}|^2 = 1.
\end{eqnarray}

Here we derive some preliminary results that are used later. 
First, Eq.~\eqref{imperfect G conditions} leads to 
$|G_{21}G_{11}^*-G_{22}G_{12}^*|^2=|G_{23}|^2|G_{13}|^2$; 
together with the other two equations, we then have 
\begin{eqnarray}
& & \hspace*{-0.6em}
\label{proof eq 1}
     \Big| \frac{G_{11}G_{22}-G_{12}G_{21}}{G_{11}} \Big|^2
\nonumber \\ & & \hspace*{2em}
      = 1 + \Big| \frac{G_{22}}{G_{11}} \Big|^2
            - \frac{|G_{12}|^2+|G_{21}|^2+1}{|G_{11}|^2}. 
\end{eqnarray}
Similarly, 
\begin{equation}
\label{proof eq 2}
     \Big| \frac{G_{13}G_{22}-G_{12}G_{23}}{G_{11}} \Big|^2
      = \Big| \frac{G_{22}}{G_{11}} \Big|^2
         - \Big| \frac{G_{23}}{G_{11}} \Big|^2 - 1. 
\end{equation}
Furthermore, we can prove that $|G_{22}/G_{11}|\rightarrow 0$ never hold 
in the limit of $|G_{11}|\rightarrow \infty$ as follows. 
If $|G_{22}/G_{11}|\rightarrow 0$, this leads to 
$G_{21}/G_{11}\rightarrow 0$ and $G_{23}/G_{11}\rightarrow 0$, 
and furthermore, $|G_{12}/G_{11}|\rightarrow 1$ and $|G_{13}|\rightarrow 1$ 
from Eqs.~\eqref{imperfect G conditions} and \eqref{proof eq 1}; 
then using Eq.~\eqref{proof eq 2} we have $G_{23}\rightarrow 0$ and 
accordingly $G_{21}G_{11}^*-G_{22}G_{12}^*\rightarrow 0$, which 
leads to a contradiction.

Now we are concerned with the amplification gain $|G_{11}^{{\rm (fb)}}|$ 
in the limit $|G_{11}|\rightarrow \infty$. 
It is given by 
\begin{eqnarray*}
& & \hspace*{-0.8em}
     |G_{11}^{\rm (fb)}|
      = \Big| \frac{1 - \alpha_1\alpha_2 K_{21}(G_{11}G_{22}-G_{12}G_{21})/G_{11}}
                             {1/G_{11} - \alpha_1\alpha_2 K_{21}(G_{22}/G_{11})} \Big|
\nonumber \\ & & \hspace*{0.25em}
      \approx \Big| \frac{-1}{\alpha_1\alpha_2 K_{21}(G_{22}/G_{11})} 
        + \frac{(G_{11}G_{22}-G_{12}G_{21})/G_{11}}{G_{22}/G_{11}} \Big|.
\end{eqnarray*}
The second term is upper bounded, because from Eq.~\eqref{proof eq 1} 
\begin{equation}
\label{proof eq 3}
     \Big| \frac{G_{11}G_{22}-G_{12}G_{21}}{G_{11}} \Big|
          \Big/ \Big| \frac{G_{22}}{G_{11}} \Big|
      \leq \sqrt{ 1 + 1\Big/ \Big| \frac{G_{22}}{G_{11}} \Big|^2}
\end{equation}
and also $|G_{22}/G_{11}|$ does not converge to zero. 
Therefore, we need $K_{21}(G_{22}/G_{11})\rightarrow 0$ (hence, 
$K_{21}\rightarrow 0$) to have the condition 
$|G_{11}^{{\rm (fb)}}|\rightarrow \infty$.

Finally, the added noise in the controlled system is computed as follows: 
\begin{eqnarray}
\label{realistic A}
& & \hspace*{-0.9em}
    {\cal A}^{\rm (fb)} = \frac{|G_{12}^{\rm (fb)}|^2 + |G_{13}^{\rm (fb)}|^2
                     + |G_{14}^{\rm (fb)}|^2 + |G_{15}^{\rm (fb)}|^2 
                     + |G_{16}^{\rm (fb)}|^2 }{2|G_{11}^{\rm (fb)}|^2}
\nonumber \\ & & \hspace*{1.5em}
     = \frac{|G_{11}^{\rm (fb)}|^2 + 2|G_{13}^{\rm (fb)}|^2 -1 }
                 {2|G_{11}^{\rm (fb)}|^2}
\nonumber \\ & & \hspace*{1.5em}
     = \frac{1}{2} - \frac{1}{2|G_{11}^{\rm (fb)}|^2}
         + \frac{|G_{13}^{\rm (fb)}|^2}{|G_{11}^{\rm (fb)}|^2}, 
\end{eqnarray}
where Eq.~\eqref{imperfect constraint} is used; 
also note that all the noise fields are now vacuum. 
The third term is given by 
\begin{equation}
\label{11_13 ratio}
\hspace{-0.56em}
     \frac{|G_{13}^{\rm (fb)}|^2}{|G_{11}^{\rm (fb)}|^2}
     = \Big| \frac{G_{13}-\alpha_1\alpha_2 K_{21} (G_{13}G_{22}-G_{12}G_{23})}
                            {G_{11}-\alpha_1\alpha_2 K_{21} (G_{11}G_{22}-G_{12}G_{21})} 
                            \Big|^2. 
\end{equation}
Then from Eq.~\eqref{proof eq 1} we find 
\[
      \Big| K_{21}\cdot\frac{G_{11}G_{22}-G_{12}G_{21}}{G_{11}} \Big|^2
      \leq  |K_{21}|^2 + \Big| K_{21}\cdot \frac{G_{22}}{G_{11}} \Big|^2
       \rightarrow 0, 
\]
in the limit $K_{21}(G_{22}/G_{11})\rightarrow 0$ and $K_{21}\rightarrow 0$. 
Also $| K_{21}(G_{13}G_{22}-G_{12}G_{23})/G_{11}|\rightarrow 0$ holds 
due to Eq.~\eqref{proof eq 2}. 
As consequence, the added noise in the limit 
$|G_{11}^{\rm (fb)}|\rightarrow \infty$ is given by 
\[
      \lim_{|G_{11}^{{\rm (fb)}}|\rightarrow \infty}{\cal A}^{\rm (fb)} 
        = \frac{1}{2} + \frac{|G_{13}|^2}{|G_{11}|^2}.
\]
Hence, together with Eq.~\eqref{realistic Ao limit}, we obtain 
Eq.~\eqref{imp added noise}.

The point of this result is that, due to the strong constraint on the noise input 
fields, which is represented by Eq.~\eqref{imperfect constraint}, the added 
noise does not explicitly contain the terms that stem from the creation input 
modes $d_4^\dagger, d_5^\dagger$, and $d_6^\dagger$. 
This is because of the passivity property of the controller 
\eqref{imperfect controller} and the feedback transmission lines 
\eqref{imperfect FB connection} that are composed of only the creation modes.


\section{Proof of Eq.~(13)}

The stability of the controlled amplifier is guaranteed if and only if all the poles 
of $G^{\rm (fb)}(s)$ lie in the left-hand complex plane. 
In our case, those are given by the solutions of the following characteristic equation: 
\[
    (1-\beta)s^2 + \kappa s + (1+\beta)\kappa^2/4 + i\beta \kappa \lambda=0.
\]
In the standard case where the coefficients of the characteristic equation are all 
real, the Routh-Hurwitz criterion can be used for the stability test, but now 
the above one contains an imaginary coefficient. 
Hence here we set $s=x+iy$, $x, y\in{\mathbb R}$, transforming the above 
equation to 
\[
       \Big[x+\frac{\kappa}{2(1-\beta)}\Big]^2 - y^2 
            = \frac{\kappa^2\beta^2}{4(1-\beta)^2},~~~
       y = \frac{-\beta \kappa \lambda}{2(1-\beta)x+\kappa}. 
\]
The poles are given by the intersections of these curves in the complex plane; 
Fig.~7 shows the case for $\beta \lambda>0$. 
Hence, for the poles to be left in the complex plane, the parameters must satisfy 
\[
       \beta\lambda < \frac{\kappa}{2}\sqrt{\frac{1+\beta}{1-\beta}}. 
\]
Considering the other case (i.e. $\beta \lambda<0$), we end up with the 
stability condition \eqref{stability condition}.

\begin{figure}[t]
\centering 
\includegraphics[width=5.7cm]{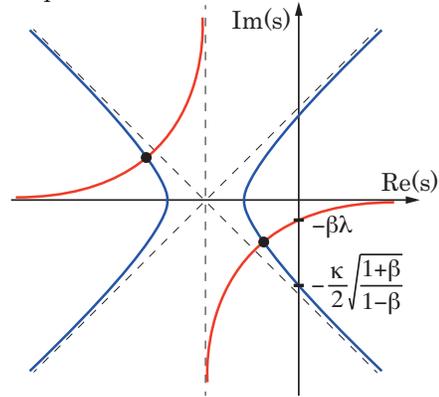}
\caption{
The poles represented by the cross points between two curves. 
}
\label{stability}
\end{figure}

Note that, as demonstrated above, in general the stability analysis becomes 
complicated for a complex-coefficient or higher-order transfer function. 
The {\it Nyquist method} \cite{Astrom Murray} is a very useful graphical 
tool that can deal with such cases, although an exact stability condition is not 
available. 
Another way is a time-domain approach based on the so-called 
{\it small-gain theorem} \cite{Zames,Doyle}, that produces a sufficient 
condition for a feedback-controlled system to be stable; 
the quantum version of this method \cite{DHelon} will be useful to test the 
stability of the controlled feedback amplifier.


\section{Phase-conjugating case}

The Hermitian conjugate of the second element of Eq.~(3) is 
given by $\tilde{b}_2=G_{21}^*b_1^\dagger+G_{22}^* b_2$. 
That is, the output $\tilde{b}_2$ is the amplified signal of the conjugated 
input $b_1^\dagger$, with gain $|G_{21}|$; 
this is called the phase-conjugating amplification. 
The feedback control in this case is almost the same as for the 
phase-preserving amplification. 
We consider the ideal feedback configuration shown in Fig.~2 (i.e., the 
noise fields $d_3,\ldots,d_6$ are ignored) and now focus on the auxiliary output 
$\tilde{b}_3=(G_{21}^{\rm (fb)})^* b_1^\dagger + (G_{22}^{\rm (fb)})^* b_4$. 
Then the amplification gain is evaluated, in the large amplification limit 
$|G_{21}|\rightarrow \infty$, as 
\begin{eqnarray*}
& & \hspace*{-1em}
      |G_{21}^{({\rm fb})}| 
          = \Big| \frac{K_{11}}{1/|G_{21}| - K_{21}G_{22}/|G_{21}|} \Big|
      \rightarrow 
       \Big|\frac{K_{11}}{-K_{21} e^{i\theta}} \Big|
\nonumber \\ & & \hspace*{2em}
      = \frac{|K_{11}|}{|K_{21}|}
      = \sqrt{\frac{1}{|K_{21}|^2}-1}. 
\end{eqnarray*}
In the first line of the above equation, we have used 
$|G_{22}/G_{21}|^2=1+1/|G_{21}|^2\rightarrow 1$; 
also the last equality comes from the unitarity of $K$, i.e., 
$|K_{11}|^2+|K_{21}|^2=1$. 
Therefore, when the original amplification gain is large ($|G_{21}|\gg 1$), 
the controlled system works as a phase-conjugating amplifier 
with gain $\sqrt{1/|K_{21}|^2-1}>1$. 
As in the phase-preserving case, this controlled gain is robust compared to 
the original one $|G_{21}|$.


\end{document}